# Optically induced dynamic nuclear spin polarisation in diamond

Jochen Scheuer[1,3], Ilai Schwartz[2,3], Qiong Chen[2,3], David Schulze-Sünninghausen[4], Patrick Carl[5], Peter Höfer[5], Alexander Retzker[6], Hitoshi Sumiya[7], Junichi Isoya[8], Burkhard Luy[4], Martin B. Plenio[2,3], Boris Naydenov[1,3*] and Fedor Jelezko[1,3]

1) Institute of Quantum Optics, Ulm University, Albert Einstein Allee 11, 89081 Ulm, Germany
2) Institute of Theoretical Physics, Ulm University, Albert Einstein Allee 11, 89069 Ulm, Germany
3) Centre for Integrated Quantum Science and Technology (IQST), Albert Einstein Allee 11, 89081 Ulm, Germany
4) Institute of Organic Chemistry and Institute for Biological Interfaces 4 – Magnetic Resonance, Karlsruhe Institute of Technology (KIT), Fritz-Haber-Weg 6, 76131 Karlsruhe, Germany
5) Bruker BioSpin GmbH, 76287 Silberstreifen, Rheinstetten, Germany
6) Racah Institute of Physics, The Hebrew University of Jerusalem, Jerusalem 91904, Givat Ram, Israel
7) Sumitomo Electric Industries Ltd., Itami, 664-001, Japan
8) Research Centre for Knowledge Communities, University of Tsukuba,Tsukuba, 305-8550, Japan

## Abstract

*The sensitivity of Magnetic Resonance Imaging (MRI) depends strongly on nuclear spin polarisation and, motivated by this observation, dynamical nuclear spin polarisation has recently been applied to enhance MRI protocols (Kurhanewicz, J., et al., Neoplasia 13, 81 (2011)). Nuclear spins associated with the $^{13}$C carbon isotope (nuclear spin I = 1/2) in diamond possess uniquely long spin lattice relaxation times (Reynhardt, E.C. and G.L. High, Prog. in Nuc. Mag. Res. Sp. 38, 37 (2011)) If they are present in diamond nanocrystals, especially when strongly polarised, they form a promising contrast agent for MRI. Current schemes for achieving nuclear polarisation, however, require cryogenic temperatures. Here we demonstrate an efficient scheme that realises optically induced $^{13}$C nuclear spin hyperpolarisation in diamond at room temperature and low ambient magnetic field. Optical pumping of a Nitrogen-Vacancy (NV) centre creates a continuously renewable electron spin polarisation which can be transferred to surrounding $^{13}$C nuclear spins. Importantly for future applications we also realise polarisation protocols that are robust againstan unknown misalignment between magnetic field and crystal axis.*

## 1. Introduction

Nuclear magnetic resonance (NMR) is one of the most powerful spectroscopic techniqueswith numerous applications in the life sciences, the material sciences and medicine. Long coherence times of nuclear spins allow to use this method for structural analysis of biomolecules and determination of molecular dynamics. Using highly sensitive NMR spectra combined with external magnetic field gradients nuclear magnetic moments can be imaged with high spatial resolution [1]. Both NMR spectroscopy and MRI critically depend on the polarisation of nuclear spins with which the observed signal scales linearly. At ambient conditions nuclear spins are very weakly polarised even in very high magnetic fields. Therefore significant effort was directed towards the development of protocols for enhancing nuclear spin polarisation above the thermal value.

---

[*]E-mail: boris.naydenov@uni-ulm.de

Dynamic nuclear polarisation (DNP) protocols that use polarisation transfer from electron spins to nuclear spins have been realised and their application in imaging has been demonstrated recently [2]. It is important to note however that polarisation transfer reaches its best performance at low temperatures where electron spins are strongly polarised already in thermal equilibrium. Furthermore, owing to the long relaxation times of electron spins at these temperatures, typically a single polarisation cycle takes about a second during which a single electron spin can polarise at most a single nuclear spin. Therefore a high concentration of polarizing paramagnetic agent is required to reach a high degree of nuclear spin polarisation. These limitations create a significant demand on novel polarisation protocols that allow for fast and efficient polarisation transfer ideally even at room temperature. Recently a novel mechanism for DNP based on optically pumped electron spins (optical dynamical nuclear spin polarisation, ODNP [3-4]) was introduced and realised in molecular crystals. However the short lifetime of the photo-excited triplet state and fast dephasing of optically active impurities at ambient conditions limit the applicability of this technique for NMR and MRI. This is the reason why the recent demonstration of diamond [5-7] as a promising ODNP platform attracted the attention of the NMR and MRI communities.

The NV centres (see Figure 1b) in diamond exhibits a long relaxation time of its electron spins even at room temperature [8] and its electronic ground state triplet can be efficiently polarised using optical excitation [5-7]. NVs can be created in bulk diamond using electron irradiation of nitrogen rich crystals or by controlled doping of ultra-pure diamonds using nitrogen ion implantation. Their wavefunction is localised to within 1 nm of these colour centres allowing them to retain thesedesirable physical properties when embedded in diamond nanocrystals with sizes below 10 nm [9]. Diamond nanoparticles doped with colour centres already found their application in fluorescence microscopy as non-bleachable fluorescence markers and drug carriers [10]. Recently nanodiamonds have been considered as promising material for applications related to DNP [5-7, 11-14]. See [15] for a recent review on biomedical applications of NV centres in nanodiamonds.

The reason why diamond is attractive for DNP and MRI applications is related to the long relaxation time of $^{13}$C nuclear spins and ability to polarise them quickly using optically pumped colourcentres. NVs can be polarised within 200 ns by a laser pulse that induces spin-selective relaxation into the $m_s=0$ sublevel of the ground state resulting in spin polarisation exceeding 96% [16].

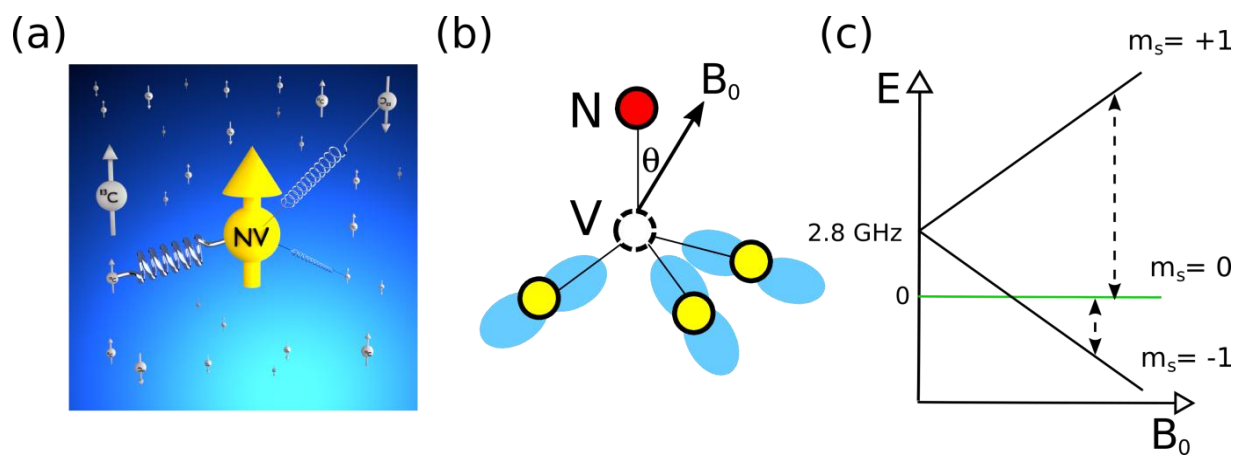

Figure 1(a) Schematic presentation of our experiments.The electron spin of the NV is polarised with a short laser pulse. Then the polarisation is transferred to the neighbouring$^{13}$C nuclei by means of a microwave induced Hartmann-Hahn resonance.The springs represent the coupling between the NV and neighbouring$^{13}$C. This process is repeated many times. (b) Crystal structure of the NV centres. The magnetic field is applied usually at θ = 0⁰ (c) Dependence of the ground state spin levels on the strength of the applied magnetic field for θ = 0⁰. The two dashed lines show the allowed ESR transitions.

In order to transfer polarisation to nuclear spins it is necessary to match the energyscale of the electron spin to that of the nuclear spin systems because the large energy mismatch between these two spins species due to the zero-field splitting (D = 2.8 GHz) of the spin triplet of NV centre prevents a direct polarisation exchange between the electron and nuclear spins. There are several methods to bringelectron and nuclear spins in resonance. The first one is to set the static magnetic field close to the level crossing (either the ground or the excited state) of the NV centre [17]. Resonance can also be achieved by microwave driving of the electron spin of the NV with a Rabi frequency that matches the nuclear spin Larmor frequency (Hartmann-Hahn condition) [18-19]. Furthermore it is possible to realise polarisation transfer by the solid effect (driving a forbiddendouble quantum transition thus inducing electron and nuclear spin flips) [5, 14]. Although the above mentioned techniques perform well for bulk diamond where NV centres have well defined orientations with respect to the static and driving fields, they do not apply directly to nanodiamond samples due to their random orientations. The strong spread of the Rabi frequencies of the microwave for chaotically oriented nanodiamonds severely reduces the probability to reach a Hartmann-Hahn resonance. Furthermore, the application of the solid effect would require an integration over a very wide range of detunings due to the large Zeeman shifts that are experienced by the NV centre. Both effects will result in a severe reduction of the total achievable nuclear spin polarisation. In this work we demonstrate the experimental realisation of a novel robust technique based on the solid effect and fast sweep of the driving field by inducing a Landau-Zener transition [14] that has been designed specifically to achieve significant nuclear spin polarisation even in the presence of random orientations of nanodiamonds.

## 2. Experiments

The experiments reported in this paper were performed on a synthetic diamond with natural abundance of $^{13}$C nuclear spins (1.1 %). The crystal was grown using the high pressure high temperature (HPHT) technique and contains 0.5 parts per million (ppm) P1 centres (a paramagnetic defect which is constituted by a single substitutional nitrogen atom). About 5 % of the P1 centres were converted to NV centres by high energy electron irradiation and subsequent annealing, achieving NV concentrationof 0.05 ppm. Figure 1a shows a sketch of the DNP experiment.

The ODNP was performed on a commercial X band Electron Paramagnetic Resonance (EPR) spectrometer (Bruker ELEXYS II) equipped with an arbitrary waveform generator and a microwave cavity allowing optical access. In order to polarise the NV centres a continuous laser beam (Laser quantum, Gem with a wavelength of 532 nm) with 3 mm diameter and 150 mW power was directed onto the crystal inside the EPR resonator. Owing to the weakNV-doping of the sample the optical density of the crystal was low (below 0.1). This allows homogeneous illumination of the sample resulting in similar polarisation conditions for the whole ensemble of NV centres. In Figure 2a the enhancement of the EPR spectrum of NVs in diamond under continuous laser irradiation is shown.

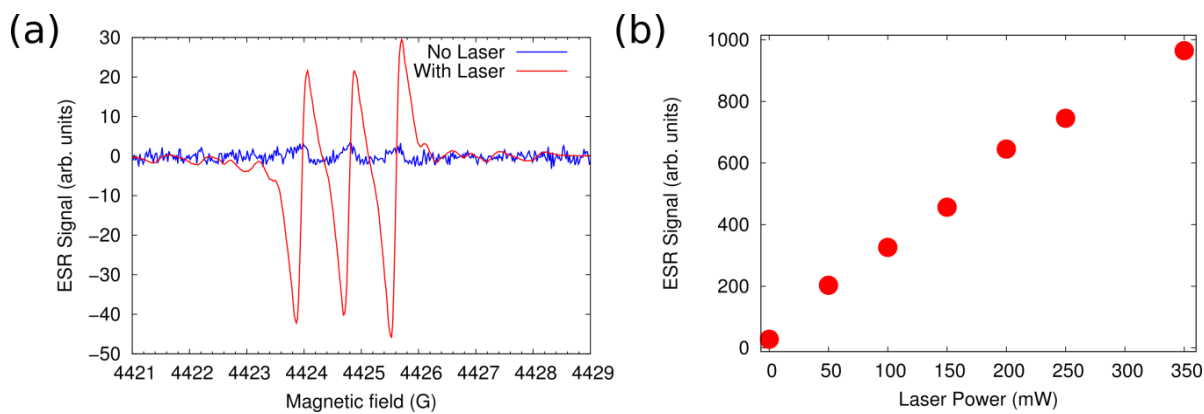

Figure 2 (a) ESR spectrum showing the NV resonance lines without (blue) and with optical pumping (red). The field was oriented along the [111] axis of the crystal coinciding with one of the four possible orientations of the NVs in the diamond. Three other orientations of NV centres have the same angle with respect to axis of NV centre and produce overlapping EPR lines. (b) Dependence of intensity of the ESR line as a function of the laser power. We could not observe saturation due to the limited laser power and to prevent heating of the sample.

As a first test of DNP we have performed polarisation transfer form NV centres to $^{13}$C using the Hartmann Hahn resonance at room temperature. This protocol, called Nuclear Spin Orientation via Electron spin locking (NOVEL) [20], brings the electron spin of the NV centre in resonance with the surrounding nuclear spins when the driving frequency of the electron spin (Rabi frequency) equals the Larmor frequency of the nuclear spins. We have previously shown that the nuclear spin polarisation resulted in narrowing of the NV's ESR line width. Such a narrowing is related to polarisation of strongly coupled nuclear spins in its surrounding ("so called "frozen core") [18]. For experiments presented in this paper we were focusing on studies of the polarisation of bulk nuclear spins which is relevant for future applications of nanodiamonds as contrast agents in NMR and MRI. Therefore the polarisation was not measured using NV centre probes, but rather by using a conventional NMR spectrometer (Bruker Fourier 300). Figure 3a shows the experimental sequence for Hartmann-Hahn DNP.

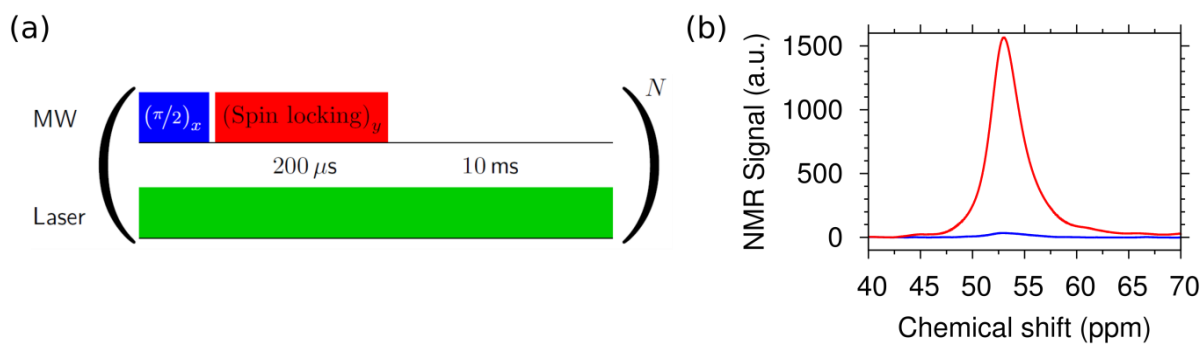

Figure 3 (a) NOVEL Pulse sequence used for hyperpolarisation of the $^{13}$C nuclear spins. The Rabi frequency of the spin locking pulse equals the $^{13}$C Larmor frequency of 4.87 MHz. (b) NMR spectra for thermally (blue) and hyperpolarised $^{13}$C nuclear spins (red). Later after a minute long probe transfer to the NMR spectrometer, an enhancement factor of about 45 is observed. Please note that for this type of samples it takes very long time for the thermal polarisation to build-up at B = 7 T, see also [21], the thermal spectrum was measured after 13h.

It consists of an electron spin locking combined with continuous illumination of the NV centres. Note that the illumination with power density of 10 W/cm$^2$ induces optical pumping at kHz rate. Therefore dephasing related to optical pumping is negligible within the duration of the spin locking pulse. The length of the latter is limited by the relaxation time of the NV electron spin in the rotating frame ($T_{1\rho}$ = 0.465 ms, data not shown). In our experiments a value of 0.2 ms was used.

A single NV centre can polarise one nuclear spin within one polarisation transfer cycle. In our crystal the concentration of $^{13}$C nuclear spins (1.1 %) was exceeding by four orders of magnitude the concentration of NV defects. Therefore in order to achieve strong nuclear spin polarisation it was necessary to apply multiple repetitions of the Hartmann-Hahn protocol. Another important condition for efficient bulk polarisation is the ability to transfer polarisation to nuclei beyond the frozen core. This can be achieved by polarizing nuclear spins at the edge of the frozen core sphere to allow for the diffusion of nuclear spin polarisation. Spin diffusion enabled by dipole-dipole interaction between the nuclear spins will then lead to polarisation transfer to nuclei in the bulk. Note that the spin diffusion is slowed down by the magnetic field gradient created by the NV centres that are in the $m_s$=+1 and $m_s$=-1 state. To ameliorate conditions for diffusion we have introduced a polarisation transfer time window during which the NV centre is initialised into the $m_s$=0 state, where there is no coupling to nuclei. During this time interval polarised nuclei in the frozen core can transfer their polarisation by diffusion to the volume outsidethe frozen core. The length of the spin locking pulse in the NOVEL sequence was chosen to reach nuclear spins with coupling constant of 10 kHz (situated approximately in a 2 nm radius around the NV centre). After five minutes of polarisation the sample was removed from the EPR spectrometer manually and transferred to the NMR spectrometer for measuring the $^{13}$C NMR signal. During the transfer,that took about a minute time across a distance of approximately 100 meters, the sample was kept in a static magnetic field (with a strength of few hundred Gauss) created by a permanent magnet. Figure 3b shows NMR spectra of non-polarised (blue line) and polarised (red line) nuclear spins. The former has been recorded after keeping the sample for 13 h in the spectrometer at a magnetic field of 7 Tesla in order to let the thermal polarisation to build-up. If the sample is measured directly after inserting it into the spectrometer no signal is observedin contrast to the polarised sample. Our hyperpolarisation procedure takes about 6 minutes, resulting in a factor of at least 130 reduction of the measurement time. Additionally we have observed a 45 fold enhancement of the signal which is in good agreement with expected polarisation enhancement (see the supplementary information for details about the simulation of the polarisation build-up).

A technique allowing to polarise nuclear spins in diamond is of particular interest for applications in MRIof diamond nanocrystals as labels. However, as stated earlier, in sampleswith random orientations, e.g. in suspensions, the Hartmann-Hahn condition is achieved only for the tiny subensemble for which the quantisation axes of NV centres are aligned with the external magnetic field. Recently, a new approach has been proposed that optimises the integrated solid effect (ISE) for randomly oriented NV centres [14].

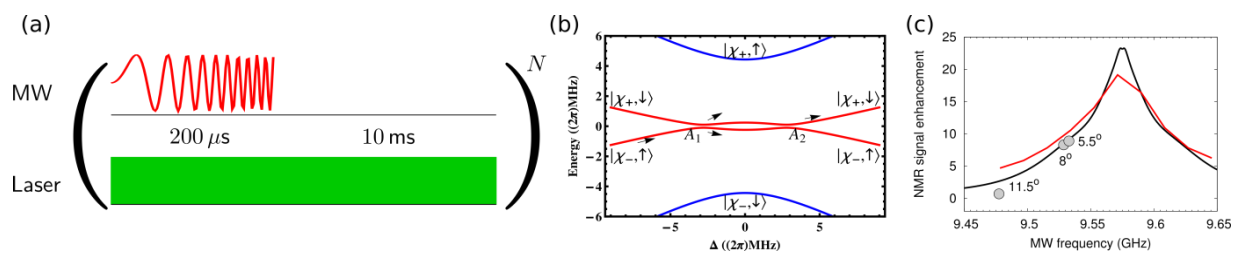

Figure 4 (a) Pulse sequence used for the ISE measurement. Instead of a spin locking sequence, the MW frequency is swept, allowing the excite transitions at different angles θ. (b) Polarisation mechanism between the NV spin and $^{13}$C nuclear spins. The solid lines depict the dependence of the eigenstates and energies on the detuning between the frequency of the microwave driving and the NV $|-1\rangle \leftrightarrow |0\rangle$ spin transition. The red lines show the states involved in the polarisation transfer, with the black arrows indicating the change of the system state in a quasi-adiabatic sweep of the microwave frequency. (c) Theoretical calculation of the nuclear spin polarisation (normalised to the thermal state) as a function of the MW frequency (black). Here we have included the bandwidth of the MW resonator (red). The circles show the enhancement of the NMR signal when we apply the ISE at different angles θ between the NV axis and the magnetic field.

This method is based on a quasi-adiabatic sweep of the driving field frequency and allows efficient population transfer for a wide range of misalignment angles (angle θ between the axis of the NV centre, see Figure *1*b, and the external magnetic field). The pulse sequence is shown in Figure 4a.

For illustrating the basic idea, let us consider a system that is composed of a NV centre and a single $^{13}$C nuclear spin in a nanodiamond of random orientation, where *θ* is the misalignment angle between the NV-axis and the magnetic field. In a strong magnetic field along the z-direction ($\gamma_e B \gg D$, with $\gamma_e B$ denoting the electron Larmor frequency, and *D* the NV centre zero-field splitting), the quantisation axis is given by the magnetic field, with the zero field splitting acting as a perturbation (see Ref. 15 and supplementary information). Focusing on the $\{|-1\rangle, |0\rangle\}$ subspace of the NV centre, we now apply a microwave driving $\sqrt{2}\Omega S_x \cos\omega_M t$ corresponding to the NV centre spin transitions $|-1\rangle \leftrightarrow |0\rangle$, ($\sqrt{2}\Omega$ is the Rabi frequency of the driving field and $\omega_M$ its frequency. In the rotating frame with the driving field frequency $\omega_M$, the Hamiltonian can be written as

$$H_{trans} = \Omega\sigma_z + \Delta\sigma_x + B_{eff}I_z + a_z\sigma_z I_z + a_x\sigma_x I_x$$

Where $\Delta = D(\theta) - \delta(\theta) - \gamma_e B + \omega_M$ is the detuning between the driving frequency and the NV $|-1\rangle \leftrightarrow |0\rangle$, spin transition ($D(\theta), \delta(\theta)$ are the angle dependent effects of the zero-field splitting, see supplementary information). $B_{eff} = \gamma_n B - A$, with $\gamma_e, \gamma_n$ being the electron and nuclear gyromagnetic ratios, and $a_z, a_x$ are the elements of the secular and pseudo-secular hyperfine interactions respectively. $\vec{I}$ is the nuclear spin-1/2 operator and $\sigma_z, \sigma_x$ are 1/2 times the Pauli matrices in the $|\pm\rangle = 1/\sqrt{2}(|0\rangle \pm |-1\rangle)$ basis.

Figure 4b shows the dependence of eigenstates of this Hamiltonian on the detuning, with $|\chi_\pm\rangle = \cos\zeta|-\rangle \pm \sin\zeta|+\rangle$ denoting the eigenstates, $\tan^{-1}\zeta/2 = \Delta/\Omega$. We note two resonance points A$_1$ and A$_2$. The polarisation is transferred between the dressed state of the electron spin and the nuclear spin during a quasi-adiabatic sweep through these resonances (see supplementary information for additional details). Note that the sweep rate should be non-adiabatic compared to the dipolar coupling (typically 50 – 200 kHz) time constant between electron and nuclear spins, i.e. around the resonance points A$_1$ and A$_2$. The upper limit on the sweep parameter is defined by the Rabi frequency of the microwave field determining the splitting between the red and blue lines in Figure 4b. Exceeding the latter would lead to the randomisation of nuclear spin polarisation.

Owing to long sample transfer time, and the limited $^{13}$C relaxation time in nanodiamonds of a few minutes [22] (we observed similar values, data not shown) compared to hours for the latter at $B = 7\ T$, our demonstrations were performed on bulk diamonds rather than in nano-diamonds. In order to realise a misalignment between NV-axis and external magnetic field, experiments were performed using different orientations of the diamond crystal with respect to the external static field but without change of the polarisation protocol. In our experimental conditions the sweep rate was chosen to be 0.3 MHz/μs and the sweep range was 100 MHz This allows to cover a misalignment angle of about 10 degrees corresponding to a~2% fraction of colour centres for isotropic orientation of NV axes, and would result in a fast build-up of polarisation in nanodiamonds, as shown by detailed numerical simulations [14]. The experiment has been performed as in the case of NOVEL. We polarise the samples for 5 min. and then transfer the diamond to the NMR spectrometer where we measure the $^{13}$C signal. The observed polarisation enhancement is demonstrated in Figure 4c. Note that slightly lower efficiency of polarisation transfer at strong misalignment angles is related to non-optimal drive at high detuning. In our experiments the microwave drive generated by an arbitrary waveform generator was kept at constant amplitude, but the microwave resonator bandwidth was limited to about 100 MHz half width at half maximum. Therefore at high rotation angle (corresponding to the resonance at the edge resonator band), the effective coupling by matching theresonance condition is much lower.

The theoretical expected polarisation enhancement was computed based on the calculated polarisation transfer probability of the ISE:

$$\bar{P} = 2P_{LZ}(1 - P_{LZ})$$

$$P_{LZ} = e^{-2\pi\mu}, \mu = \frac{\Omega^2 a_{x'}^2}{16v(\gamma_n B)\sqrt{(\gamma_n B)^2 - \Omega^2}}$$

With $\Omega$ denoting the Rabi frequency, $a_{x'}$ the effective dipolar coupling, $v$ the frequency sweep rate, and $\gamma_n B$ the nuclear Larmor frequency. The polarisation rate, which depends on the cycling time, is then averaged over an ensemblefor a nuclear spin bath of 500 spins around the NV centre, and a steady-state polarisation depending on the $^{13}$C relaxation time in the EPR magnetic field (see supp. mat.). Nuclear spin diffusion is limited by the dipolar coupling between the $^{13}$C spins (2 kHz for next neighbours) and it does not seem to be a bottleneck of the polarisation build-up in our experiments, due to the 10 ms polarisation transfer time between cycles with the NV in the $m_s$=0 state.

### 3. Conclusions

Our results show that optical nuclear spin polarisation in diamond can be performed for a wide range of orientation of NV centres at room temperature. Our protocol based on quasi-adiabatic sweep of microwave field across the resonance points allows to polarise diamond nanoparticles. Such functionalised diamond nanocrystals can serve as agent for molecular imaging using MRI techniques. Our method can be in principle transferred to other nanoparticles containing electron spins that can be optically polarised.

**Acknowledgements**:

This work was supported by the ERC Synergy grants BioQ and PoC, the EU (DIADEMS, EQUAM, SIQS), the DFG (SFB TRR/21), Volkswagenstiftung, JSPS KAKENHI (No.26246001) Bundesministerium für Bildung und Forschung (ARHES award) and an Alexander von Humboldt Professorship. BN is grateful to the Postdoc Network program of the IQST. IS acknowledges a PhD fellowship of the IQST.